\documentclass[floatfix,10pt,onecolumn,showpacs,amsmath,amssymb]{revtex4}

\usepackage{epsf}
\usepackage{graphicx}  
\usepackage{dcolumn}   
\usepackage{bm}        

\newcommand{\be}{\begin{equation}}
\newcommand{\en}{\end{equation}}
 \newcommand{\bea}{\begin{eqnarray}}
 \newcommand{\ena}{\end{eqnarray}}

\begin{document}

\title{Ghost free massive gravity with singular reference metrics}
\author{Hongsheng Zhang$^{1,2~}$\footnote{Electronic address: hongsheng@shnu.edu.cn},  Xin-Zhou Li$^1$ \footnote{Electronic address: kychz@shnu.edu.cn} }
\affiliation{ $^1$Center for
Astrophysics, Shanghai Normal University, 100 Guilin Road,
Shanghai 200234, China\\
$^2$State Key Laboratory of Theoretical Physics, Institute of Theoretical Physics, Chinese Academy of Sciences, Beijing, 100190, China
}


\begin{abstract}
An auxiliary metric (reference metric) is inevitable in massive gravity theory. In the scenario of gauge/gravity duality, a singular reference metric corresponds to momentum dissipations, which describes the electric and heat conductivity for normal conductors. We demonstrate in detail that the massive gravity with singular reference metric is ghost-free.

\end{abstract}


 \maketitle

\section{Introduction}

The translational invariance is a fundamental symmetry in the studies of physics, but it is usually broken in occurrence of realistic matters. The breaking translational invariance ought to engender some significant consequences for real systems. For nearly independent particles the consequences have good interpretations. For instance, quantum random walk yields strong localization, and the energy band is the results of the periodicity of the space (lattice structure). However, our understandings are less for the results of translational symmetry broken in strongly interacting systems. The recent developments of gauge/gravity duality provides a powerful toolkit to resolve  this problem. Because there is no momentum dissipations in the dual gauge theory of a gravity with diffeomophism invariance, Eintein gravity is a proper tool to study superconductors. Generally, to describe the properties of normal conductors requires a gravity theory in which the diffeomophism invariance is broken. More specifically, we need a gravity without spatial translational symmetry.  The massive gravity with a special singular reference metric can break the spatial translational symmetry at the boundary, and thus is helpful to study the physics of the strongly interacting system with translational symmetry broken.

In Einstein gravity, there are two massless propagating modes. After quantization, these two modes become two types of massless gravitons with different polarizations.  Theoretically, one may always be curious that what will happen if we introduce a mass for the graviton? A first try along this road is the Fierz-Pauli  massive gravity \cite{FP1}, in which the graviton is successfully endorsed a mass in linear level. However, a massive spin-2 field propagates five degrees of freedom no matter how tiny its mass is. This subtlety yields the essence of the van Dam, Veltman and Zakharov (vDVZ) discontinuity \cite{van}. This puzzle was resolved by the Vainshtein mechanism \cite{vainshtein}. Due to the strong couplings, the scalar interaction is shielded and Einstein gravity can be recovered. But the ghost freedom, call  Boulware-Deser ghost, will reappear when one extend the Fierz-Pauli massive gravity to non-linear level \cite{BDghost}. In 2010, a non-linear ghost-free massive gravity is proposed in \cite{dRGT}, in which the reference metric is set to be a Minkowskian one. The ghost problem with a general  full rank reference metric is investigated in \cite{hassan}. Several aspects of massive gravity has been studied, for example the Misner-Sharp mass and unified first law \cite{self1}, based on some previous studies \cite{self2}. The Drude peak of the normal conductors is investigated in the scenario of gauge/gravity duality in \cite{vegh}, in which a massive gravity with singular reference metric is involved. Then momentum dissipations as a duality of massive gravities with singular reference metrics are investigated in several cases\cite{many}. However, the stability of a massive gravity with singular reference metric has not been thoroughly studied, through it is mentioned in some works, for example \cite{vegh, self1}. In this paper, we present a systematic treatment of this problem.

This paper is organized as follows. First, we give a short review of the ADM formulation of Einstein gravity and massive gravity, concentrating on the ghost problem in section II. In section III, we demonstrate that the massive gravity with singular reference metrics is ghost-free. In section IV, we present our conclusions and make a brief discussion about our results.

\section{ghost problem in massive gravity}

 Before investigating the ghost problem in the massive gravity, we first make a concise review of the ADM formulation of Einstein gravity. In general relativity a standard ADM decomposition of the metric reads
  \be
 ds^2=-(N^2-N_iN^i)dt^2+2N_idtdx^i+\gamma_{ij}dx^idx^j,
 \en
 where $N$ denotes the lapse and $N_i$ denote the shift functions, and $\gamma_{ij}$ is the spatial metric. The contravariant form of the metric reads,
  \be
  g^{ab}=\frac{1}{N^2}\left[-\left(\frac{\partial}{\partial t}\right)^a\left(\frac{\partial}{\partial t}\right)^b+ \left(\frac{\partial}{\partial t}\right)^aN^i\left(\frac{\partial}{\partial x^i}\right)^b+\left(\frac{\partial}{\partial t}\right)^bN^i\left(\frac{\partial}{\partial x^i}\right)^a +(N^2\gamma^{ij}-N^iN^j)\left(\frac{\partial}{\partial x^i}\right)^a\left(\frac{\partial}{\partial x^j}\right)^b\right].
  \label{upg}
  \en

  Under such a standard decomposition, the Hilbert-Einstein action becomes,
 \be
 S=\frac{1}{2\kappa^2}\int d^4x (\pi^{ij}\dot{\gamma}_{ij}+N_{\mu}R^{\mu}),
 \label{actionadm}
 \en
 in which $(\pi^{ij}, \gamma_{ij})$ are conjugate pairs with respective to this action, $\kappa$ is the inverce of Planck mass $m_{pl}$, $\gamma_{ij}$ is the spatial metric, $N_{\mu}=(N,~N_i)$, and
 \be
 R^0=\sqrt{\gamma} \left[\textbf{R}+\frac{1}{\gamma}(\frac{\pi^2}{2}-\pi_{ij}\pi^{ij})\right],
 \label{hami}
 \en
 \be
 R^i=2\sqrt{\gamma}~\nabla_j\left(\frac{\pi^{ij}}{\sqrt{\gamma}}\right),
 \label{shami}
 \en
  where $\gamma$ denotes the determinant of $\gamma_{ij}$,  $\textbf{R}$ is the three dimensional Ricci scalar yielded by $\gamma_{ij}$. To compare with the case of massive gravity, now we present a short analysis about the freedoms in the field equation. Because the spatial metric $\gamma_{ij}$ is a symmetric one, there are six propagating degree of possible freedoms at the most, among which three freedoms are ordinary translational degree of freedoms of a mass point in three directions, two freedoms are polarization degree of freedoms, and the residue one may become a ghost, which is something like a  longitudinal photon with false kinetic term. However, there are four lagrange multipliers $N_{\mu}$ in the action (\ref{actionadm}), which impose four constraints in the resulting equations of motions. Thus, there are only two polarization degree of freedoms really $free$ in general relativity. From another point of view, we have twelve (six pairs) of dynamical variables, and eight constraint equations (four from $N_{\mu}$ and another four are Bianchi identities). The residue four dynamical variables describe two propagating degree of freedoms.

  The situation is a bit complicated in the massive case. First we discuss the minimal massive gravity \cite{revdRGT}. The stability property of the general covariant massive gravity is similar to the minimal one.  The action of the minimal massive gravity reads,
  \be
  S=\frac{1}{2\kappa^2}\int d^4x \left(\pi^{ij}\dot{\gamma}_{ij}+N_{\mu}R^{\mu}+m^2V(N_{\mu},\gamma_{ij},f)\right),
   \label{actionmass}
    \en
    where
    \be
    V=-2N\sqrt{\gamma}({\rm tr} \sqrt{I^{\mu}_{~~\alpha}}-A), ~~~~~I^{\mu}_{~~\alpha}=g^{\mu\nu}f_{\nu\alpha}.
    \en
    Here, $f$ is the reference metric. We concentrate on the case of a singular reference metric in this article. Because the cosmological term has no affect on the stability property of the massive gravity, we omit it in what follows in this paper.  The radical sign is to extract the square root of the matrix $I^{\mu}_{~~\alpha}$. Generally, an $n\times n$ matrix has $2^n$ square roots. For example the matrix diag(1,1) has four square roots: diag(1,1), diag(1,-1), diag(-1,1), and diag(-1,-1). So the operation of square root for a matrix is alive with ambiguity. We take the matrix with Lorentzian signature as the proper square root of $I^{\mu}_{~~\alpha}$. One is easy to confirm that only this choice can reduce to the  Fierz-Pauli linear massive gravity when $g_{\mu\nu}$ degenerates to a Minkowski metric. The term $A\equiv$rank($g_{\mu\nu}$)$-1$ is to ensure the term $V$ does not include a cosmological constant. For a full rank $g_{\mu\nu}$, $A=3$.


       Apparently, all $N_{\mu}$ are no longer lagrange multipliers, and thus the constraints (\ref{hami}) and (\ref{shami}) are turned off.
       So all the six possible degrees of freedom of $\gamma_{ij}$ are liberated, including the ghost. The equations of motion of $R^{\mu}$ read,
       \be
       R^{\mu}+m^2\frac{\partial V}{\partial N_{\mu}}=0.
       \label{pn}
       \en
        Thus, it is clear that $R^{\mu}$ are no longer constraints. This is the argument from Boulware and Deser, who claim that general non-linear massive gravity will be plagued by ghosts.  Note that the Bianchi identities are able to reduce the dynamical freedoms only when they work with the constraint equations. Now the constraints vanish, therefore the Bianchi identities "can not clap with one palm", doing no work to reduce the dynamical freedoms.

        \section{ghost-free massive gravity with a singular reference metric}

        The unavoidability of the  ghost is a facade of massive gravity. We make the following arguments to find the necessary condition to kill ghost. And one will see it is also a sufficient condition. To kill the ghost, we need to recover a Hamiltonian constraint in an ADM decomposition. In a reparameterized ADM decomposition, a recovered Hamiltonian constraint requires,
        \be
        F(R^{\mu})=0,
        \label{con1}
        \en
        in any decomposition, including the present decomposition. Here $F$ is a function of $R^\mu$. This constraint leads to a constraint from (\ref{pn}),
        \be
        F(\frac{\partial V}{\partial N_{\mu}})=0.
        \en
        Regarding the spatial metric $\gamma_{ij}$ and the reference metric $f$ as parameters, the above equation is essentially a constraint of $N_\mu$,
        \be
        C(N,N_i)=0.
        \en
        Hence, the ghost-free condition implies that there are only three independent variables in the four dimensional parameter space $(N, N_i)$, which expand a three dimensional hypersurface. We introduce the intrinsic coordinates $(M_1,~M_2,~M_3)$ for this hypersurface. Using the intrinsic coordinates the equations of this hypersurface read,
        \be
        M_i=M_i(N, N_j).
        \en
        Inversely, we have
        \be
        N_j=N_j(M_i,N).
        \label{NJ}
        \en
        Here, we treat $N$ as a parameter in this coordinates transformation. To recover the Hamiltonian constraint under the new parametrization of AMD decomposition, the action form under this new parametrization should be,
        \be
        S=\frac{1}{2\kappa^2}\int d^4x NF(R^\mu)+...~,
                \en
       and $N$ does not appear besides the term $NF(R^\mu)$ (and thus we obtain the constraint (\ref{con1})). So $N_j$ have to be linear to $N$ in the transformation (\ref{NJ}), otherwise the higher order terms of $N$ will appear after transformation from the second term in (\ref{actionmass}). One can confirm that the Fierz-Pauli  massive gravity can be written in this form.  According to this requirement we set
  \be
  N^i=P^i(M^k,\gamma_{kj})+NQ^{i}(M^k,\gamma_{kj}).
  \label{repara}
  \en
  We emphasize that $P^i$ and $Q^i$ are not functions of $N$.

  Then we work out the square root of $I^\mu_{~~\nu}$. From (\ref{upg}), we have
  \be
  N^2I^\mu_{~~\nu}= \begin{bmatrix} -f_{00}+N^if_{i0} & -f_{0j}+N^if_{ij} \\ N^2\gamma^{ij}f_{j0}+N^if_{00}-N^iN^jf_{j0} & N^2\gamma^{ik}f_{kj}+N^if_{0j}-N^iN^kf_{kj} \end{bmatrix}.
  \en
  We primarily concentrate on a singular reference metric, without restricted to the singular metric specially for momentum dissipations in gauge/gravity duality.
  To warm up, we consider the simplest case, in which there is only one non-zero component, assuming to be $f_{00}$. In this case, it is easy to obtain $A=0$. Further, we have,
  \be
   N(I^\mu_{~~\nu})^{\frac{1}{2}}=\sqrt{-f_{00}}\begin{bmatrix} 1 & \bf{0_1}\\-N^i & \bf{0_2} \end{bmatrix},
   \en
  where $\bf{0_1}$ is a row zero-vector and $\bf{0_2}$ is a 3 order square zero-matrix,
  and thus,
  \be
  V=-2\sqrt{\gamma}\sqrt{-f_{00}}.
  \en
  The massive term is not a function of $N$. Thus the Hamiltonian constraint is preserved. Under this situation we need not make any transformation to highlight the ghost-free property. Then we consider a dual case, in which we set the reference metric $f_{\mu\nu}$,
    \be
    f_{\mu\nu}=\begin{bmatrix} 0 & \bf{0_1} \\ \bf{0_3} & f_{ij} \end{bmatrix},
  \label{rank3}
  \en
 where $\bf{0_3}$ is a column vector. In this case, it is easy to obtain $A=2$.
 Now we set
 \be
 c^i_j=M^iM^kf_{kj},
 \label{firstp}
 \en
 and
 \be
 L^i_{~j}=\rm{tr}( {c^i_{~j}})\delta^i_{~j}+c^i_{~j}.
 \en
 Treating $L$ as a matrix, we define its inverse operator,
 \be
 L^{-1}=\left( \rm{tr}( {c^i_{~j}})\textsf{E}+MM^Tf\right)^{-1},
 \en
 where $\textsf{E}$ is a $3\times 3$ unit matrix, $M$ is a column vector of $M^i$, $M^T$ is the transposition of this column vector, and $f$ denotes the matrix of $f_{ij}$, which satisfies
 \be
 (L^{-1})^i_{~j}L^j_{~k}=\delta^i_{~k}.
 \en
 Note that the result in \cite{hassan} cannot be applied here. Then we define,
 \be
 D^i_{~j}=\sqrt{\gamma^{ij}f_{jm}L^m_{~n}}~\left(L^{-1}\right)^n_{~j},
 \en
in which we take a positive definite matrix for the result in the radical sign.
 The transformation equation (\ref{repara}) becomes
 \be
  N^i=M^i+ND^i_{~k}M^k.
  \en
 After such a transformation, the action becomes
 \be
  S=\frac{1}{2\kappa^2}\int d^4x \left[\pi^{ij}\dot{\gamma}_{ij}+N\left(R^0-2m^2\sqrt{\gamma (\rm{tr} c^i_{~j})}~ \rm{tr} D^{i}_{~j}-2\right) +M_{i}R^{i}-2m^2\sqrt{\gamma (\rm{tr} c^i_{~j})}\right],
   \label{actionmassp}
    \en
  which yields the Hamiltonian constraint under the new ADM parametrization,
  \be
   R^0-2m^2\sqrt{\gamma (\rm{tr} c^i_{~j})}~ \rm{tr} D^{i}_{~j}-2=0.
   \label{moha}
   \en
   Apparently, it is still not in the required form (\ref{con1}). Recalling that the equations corresponding to the shifts can be solved by the deformed constraint equations (\ref{pn}), we can explicitly show that (\ref{moha}) has an exact form of (\ref{con1}). With the new shifts $M_i$, the deformed constraint equation (\ref{pn}) becomes,
   \be
   R^{i}+m^2\frac{\partial V}{\partial M_{i}}=0.
       \label{pn1}
       \en
   In principle the solutions of the above equation can be written as,
   \be
   M_{i}=M_{i}(R^{j}).
   \en
   And the matrix $D$ is a function of $M^i$,
   \be
   D^i_{~j}=D^i_{~j}(M_k).
   \label{last}
   \en
  Then, finally we reach that equations (\ref{pn1}) imply  $D^i_{~j}=D^i_{~j}(R^{k})$ and equation (\ref{moha}) is the desired (modified) Hamiltonian constraint. This constraint kills the ghost excitation. More exactly, this constraint working with the zero-component of the Bianchi identity can suppress the pair of the dynamical variable of the ghost excitation.

  The next case which we will considered is the original case motivated by the momentum dissipation in the normal conductor in the scenario of gauge/gravity dualityity, in which the reference metric $f$ reads,
  \be
    f_{\mu\nu}=\begin{bmatrix} \bf{0_{2\times 2}} & \bf{0_{2\times 2}} \\ \bf{0_{2\times 2}} & f_{mn} \end{bmatrix},
  \label{rank2}
  \en
  where $\bf{0_{2\times 2}}$ is a ${2\times 2}$ zero-matrix. This case is similar to the previous one. All the equations from (\ref{firstp}) to (\ref{last}) can be applied in this case if we reset the running interval from (1,2,3) to (2,3) of the Latin indexes. Thus we proved that the theory is also ghost-free in this case.

  In a word, we demonstrate that the four dimensional minimal massive gravity with any singular reference metric is ghost-free.

 \section{conclusion and discussion}
 The massive gravity gets significant approaches recently. Generally a massive gravity has six propagating modes. Five of them are physical modes, and the residue one may be a ghost. In a class of covariant massive gravity, the ghost is shown to be merely a false appearance. Through a proper reparametrization of the ADM decomposition, one can show that the ghost excitation can be suppressed by a recovered Hamiltonian constraint. However, the applicable range this discussion is limited to a massive gravity with a full rank reference metric. We emphasize that the result in \cite{hassan} can not be directly applied in the case of singular reference metrics.  The massive gravity with a singular reference metric gains important applications to describe the momentum dissipations in several condensed matters systems, such as the electric and heat conductivity in normal conductors in the scenario of gauge/gravity duality. So it is necessary   to present a sound proof of the ghost-free property for the massive gravity with a singular reference metric.

 In this paper we made detailed demonstration that the massive gravity with a singular reference metric is ghost-free. First we show that the theory with rank-1 reference metric is ghost free without any transformations in the lapse-shift space. The Hamiltonian constraint is preserved automatically. Then we find the proper transformations in the lapse-shift space for a rank-3 reference metric, which can clearly illustrate the Hamiltonian constraint. The case of a rank-2 reference metric is similar to the case of rank-3 metric. So, we prove that the massive gravity is ghost-free with any cases of singular reference metrics. A question will come very naturally from this result. Can we recover more constraints other than the Hamiltonian constraint via transformations in the lapse-shift space? The answer is no. Because the residue five degree of freedoms are all physical freedoms. For example, if a momentum constraint is recovered, one will see that the graviton travels on this this direction with zero-velocity or a constant velocity. It is impossible for a massive graviton.

\section{Acknowledgements}
This work is supported by the Program for Professor of Special Appointment (Eastern Scholar) at Shanghai Institutions of Higher Learning, National Education Foundation of China under grant No. 200931271104, and National Natural Science Foundation of China under Grant No. 11075106 and 11275128.

\end{document}